# Rejection mechanisms for contaminants in polymeric reverse osmosis membranes (pre-print Aug. 10, 2015; see also J. Membrane Science 509:36-47, 2016 http://dx.doi.org/10.1016/j.memsci.2016.02.043 )

Meng Shen,[1] Sinan Keten,[1,2] Richard M. Lueptow[1,3]

[1] *Mechanical Engineering, Northwestern University, Evanston, IL, USA*

[2] *Civil & Environmental Engineering, Northwestern University, Evanston, IL, USA*

[3] *The Northwestern Institute on Complex Systems (NICO), Northwestern University, Evanston, Illinois, USA*

**Abstract**

Despite the success of reverse osmosis (RO) for water purification, the molecular-level physico-chemical processes of contaminant rejection are not well understood. Here we carry out non-equilibrium molecular dynamics (NEMD) simulations on a model RO membrane to understand the mechanisms of transport and rejection of both ionic and inorganic contaminants in water. While it is commonly presumed that the contaminant rejection rate is correlated with the dehydrated solute size, this approximation does not hold for the organic solutes and ions studied. In particular, the rejection of urea (2.4 Å radius) is higher than ethanol (2.6 Å radius), and the rejections for organic solutes (2.2-2.8 Å radius) are lower than $Na^+$ (1.4 Å radius) or $Cl^-$ (2.3 Å radius). We show that this can be explained in terms of the solute accessible free volume in the membrane and the solute-water pair interaction energy. If the smallest open spaces in the membrane's molecular structure are all larger than the solute size including its hydration shell, then the solute-water pair interaction energy does not matter. However, when the open spaces in the polymeric structure are such that solutes have to shed at least one water molecule to pass through a portion of the membrane molecular structure, as occurs in RO membranes, the pair interaction energy governs solute rejection. The high pair interaction energy for water molecules in the solvation shell for ions makes the water molecules difficult to shed, thus enhancing the rejection of ions. On the other hand, the organic solute-water interaction energies are governed by the water molecules that are hydrogen bonded to the solute. While these hydrogen bonds have pair interaction energies that are much larger than that of the non-hydrogen bonded water molecules in the solute solvation shell, they are significantly less the ion-water pair interaction energy. Thus, organic solutes more easily shed water molecules than ions to pass through the RO membrane. Since urea molecules have more hydrogen-bonding sites than alcohol molecules, urea sheds more hydrogen-bonded water molecules and forms more hydrogen bonds with the membrane leading to a higher rejection of urea than occurs for ethanol, a molecule of similar size but with fewer hydrogen bonding sites. These findings underline the importance of the solute's solvation shell and solute-water-membrane chemistry in the context of reverse osmosis, thus providing new insights into solute transport and rejection in RO membranes.

## 1. Introduction

Reverse osmosis (RO) membrane separation is a widely used technology for drinking water purification [1] because it is energy efficient, requiring no high temperatures or phase transitions [2]. The performance of practical polymeric RO membranes can be predicted using continuum-level models such as the solution-diffusion model [3]. However, the molecular level physico-chemical transport and rejection mechanisms in RO membrane are still not clear. Molecular dynamics (MD) simulations with properly validated inter-atomic force fields offer the ability to study these mechanisms, because the movement of individual solute, solvent, and membrane molecules can be easily tracked, and chemical interactions such as hydrogen bonding can be isolated.

In RO membrane separation, a high pressure is applied to the contaminated feed solution on one side of the membrane that opposes the osmotic pressure. In the ideal situation, water molecules pass through the membrane against the osmotic pressure, but contaminant solute ions and molecules do not pass. Although the rejection of monovalent ions by typical RO membranes is often greater than 98% [4], the rejection of small organic solutes, some of which have larger dehydrated molecular sizes than $Na^+$ and $Cl^-$, can be less than 80% [5]. Small organic compounds, which are potentially harmful to human health, can be found in produced water in the oil and natural gas industry as well as other situations [6]. Of course, it is also necessary to remove organic compounds from drinking water. However, it is difficult to experimentally investigate the underlying mechanisms that lead to high rejection of small charged ions but lower rejection of larger neutral organic solutes. Molecular dynamics simulations offer a means to directly consider the movement of individual ions and molecules as they approach and transit through the membrane. For example, the importance of water molecules forming a "shell" surrounding ionic contaminants has been investigated using molecular dynamics in narrow pores [7], but the surrounding "water shell" and other aspects of the chemistry of organic solute transport in RO membranes has not been studied.

Equilibrium molecular dynamics simulations have previously been used to study water and ion transport in polymeric reverse osmosis membranes [8-17]. Kotelyanskii et al. observed a "jump-like" movements for water within the membrane [9] and a lower mobility of $Cl^-$ than $Na^+$ in a hydrated membrane [8]; Harder et al. developed a heuristic method to computationally construct the polymeric RO membrane and considered water transport through the membrane [10]; Luo et al. predicted water flux and ion rejection by calculating diffusivity and the free energy landscape from molecular dynamics under equilibrium conditions based on a solubility-diffusion theory [11]; Hughes et al. studied the free energy surface associated with selected ion contaminant/molecular foulant pathways using umbrella sampling methods [16, 17]; Ding et al. considered the structure and dynamics of water molecules confined in polymeric membranes [12, 13]; Kolev et al. simulated a membrane that closely matches the known physical properties of commercial membranes [14]; Drazevic et al. assessed the validity of hindered transport theory to RO membranes by experiments and molecular dynamics simulations [15].

While significant progress has been made in understanding some aspects of the physico-chemical processes in RO molecular dynamics studies as described above, all have been under equilibrium conditions. Actual RO membranes function only under non-equilibrium conditions where there is a large transmembrane pressure that opposes and exceeds the osmotic pressure. Non-equilibrium conditions have been introduced in molecular dynamics to study water/ion transport in inorganic and biological membranes [18-20]. However, only very recently have non-equilibrium molecular dynamics simulations been used to investigate water transport and contaminant rejection in polymeric RO membranes [21-23]. In this approach the actual operating conditions of the RO process are simulated by applying a pressure difference across the membrane. Progress has been achieved in understanding some aspects of water and ion transport, though two recent studies seem to consider the case of equal ion concentrations on both sides of the membrane [21, 22], a situation that would not occur in practice. By contrast, we study the situation in which solutes are initially concentrated on the high-pressure feed side of the membrane with pure water on the low-pressure permeate side of the membrane, both in our recent work in which we considered the physical aspects of the transport including the impact of the percolated free volume and size of the solute [23] and the current study. Furthermore, in contrast to these recent non-equilibrium studies that only considered water permeability and ion rejection [21, 22], we are able to more deeply probe the chemistry of solute transport and rejection by considering the transport of two ions, $Na^+$ and $Cl^-$, as well as four organic solutes, methanol, ethanol, 2-propanol, and urea. This choice of solutes along with using non-equilibrium simulation conditions provides the opportunity to thoroughly examine not only the impact of solute size, but also the comparative consequences of mechanisms such as solute hydration and hydrogen bonding on solute transport and rejection in polyamide RO membranes.

**2. Methodology**

The non-equilibrium molecular dynamics (NEMD) simulation approach is described in detail in our previous publication [23], so we only provide highlights here. A cross-linked polymeric FT-30 membrane was computationally constructed using a heuristic method [10] wherein m-phenylene diamine (MPD) and trimesoyl chloride (TMC) monomers were assembled randomly in a computational box to form a polymeric membrane structure. The membrane was computationally hydrated by filling in the open space between membrane polymer chains with water molecules. Then the hydrated membrane was then placed between a feed reservoir of an aqueous solution with 192 solute molecules or ion pairs and 5000 water molecules on the left, corresponding to a solute concentration of about 2.13 M, and a permeate reservoir of pure water with 5000 water molecules on the right. Graphene sheets or rigid sheets of water molecules were added to the free surfaces of the feed solution reservoir on the left and pure water reservoir on the right. A pressure difference across the membrane was introduced by applying forces to each atom in the graphene/water sheets with the force on the left greater than that on the right thereby generating a non-equilibrium pressure difference across the membrane. Periodic boundary conditions were used in the depthwise x- and vertical y- directions with the computational domain extending approximately 55.1 Å in the x-direction and 53.8 Å in the y-direction. Unlike previous studies where there was only one reservoir [8, 11], this configuration mimics the actual filtration process with reservoirs on both sides of the membrane. To allow vibrations of most membrane atoms while fixing the membrane in the z-direction, which is perpendicular to the plane of the membrane, a small number of atoms of the membrane were pinned to fixed points in space during the simulations.

Before the pressure-driven transport simulations commenced, forces corresponding to pressures of 0.1 MPa were applied to the liquid on both sides, and the simulations were run for about 10 ns to saturate the membrane with water. The densities of the dry and hydrated membranes are close to previous experimental and simulation values [11, 14, 16, 24, 25]. Like other recent studies [21, 22], the simulated membranes under study are much thinner than actual polymeric membranes used commercially, which are typically about 0.2 µm thick [26]. Nevertheless, the water flux through the simulated membrane is in the range expected for commercial membranes [23]. More importantly, the atomic level structure of the simulated membrane is adequate to provide an opportunity to investigate the effects of local membrane structure on water transport and solute rejection.

Four organic solutes, methanol, ethanol, 2-propanol and urea, and sodium chloride, all of which are soluble in water, were considered in the simulations. Details of the molecular models and parameters for the solutes are provided in our previous publication [23]. In the NEMD simulations of pressure-driven transport, pressures on the solution side was 150 MPa, while the pressure on the pure water side was maintained at 0.1 MPa. The simulations were run long enough to obtain about 23 ns of data.

It is obvious that the small size of the membrane (a few nm) and short duration of the simulations (tens of nanoseconds) make statistically significant measures of macroscale properties of the membrane difficult. Therefore, we focus on comparative results to obtain a better understanding of the physico-chemical mechanisms at play in reverse osmosis. That is, we consider similarities and differences between the four organic solutes and two ionic solutes in order to evaluate the underlying mechanisms of transport and rejection. While full scale simulations are well beyond the reasonable capabilities of existing computational resources, our approach provides substantial insight into molecular level flux, transport, and rejection mechanisms at a reasonable computational cost.

## 3. Results

### 3.1. Solute transport: trajectories, path length and passing time

Consider first a comparison of the trajectories of the water and solute molecules that pass through the membrane. As shown in our previous work, the solute trajectories are quite random as solute molecules collide with water molecules and membrane polymer chains [23]. This is evident in Fig. 1(a) in which the individual dots indicate a single solute molecule's positions every 2 ps for three 2-propanol molecule trajectories that pass through the entire thickness of the membrane, each shifted so that they do not overlap for clarity. It is evident that the solutes transit from one void, a region dense with dots indicating a portion of the membrane where the molecules spends significant time presumably because it is less densely packed with molecular chains of the membrane, to another void, consistent with our previous results [23] and those for water [22]. A curve is fitted to the data points using Savitzky-Golay algorithm [27], which allows us to extract trajectory information from the noisy data without missing local features of the trajectory. The transparent blue panels represent the boundaries of the dense region of the membrane based on the concept of a Gibbs Dividing Surface [28], and the dark blue circles on the transparent panels indicate where the 2-propanol molecules pass through this surface.

Repeating the above procedure to follow many water or solute molecules allows us to visualize trajectories that the molecules take as they transit the membrane. The trajectories for water molecules shown in Fig. 1(b) are quite dense in the membrane, indicating the many percolated transport pathways that are available for water molecules. In fact, only about 10% of the water molecule trajectories are shown in the main plot in Fig. 1(b) for clarity. The inset in Fig. 1(b)) which includes all of the trajectories for water molecules in a small region of the membrane further demonstrates the density of the trajectories for water. Although it is not readily evident in Fig. 1(b), the entry and exit points of the water molecule trajectories are quite dense in some areas of the membrane boundary but sparse in other areas. This suggests the inhomogeneity of the membrane structure with preferred pathways for the water molecules. It is also evident that the water molecules seem to transit the membrane via several primary routes where the trajectories are denser. Preferred transit routes in RO membranes were also observed previously [11, 16]. These pathways likely correspond to regions of high percolated free volume in the membrane [23].

The trajectories for all four organic solutes are shown in Fig. 1(c)-(f). ($Na^+$ and $Cl^-$ ions did not pass through the membrane during the simulations and therefore could not be included in Fig. 1.) The four organic solutes tend to transit through the same regions within the membrane volume, a result that is particularly evident for methanol and ethanol. In fact, these three regions correspond to the regions that are preferred for water transport. Though these common pathways through the membrane might be thought of as individual pores, they are much more complex and smaller than regular cylindrical or tubular pores assumed in the pore-flow models [29]. The pathway labeled A is common for all four organic solutes and water, but the lower two pathways were not used by 2-propanol molecules, at least within the constraints of the duration of the simulation for which only a limited number of solute molecules pass through the membrane. (Note that the three trajectories evident in Fig. 1(e) for 2-propanol are all along the same pathway but are shown offset from one another for clarity in Fig. 1(a).) It is also evident, by comparing Fig. 1(b)-(f), that the molecule size plays a significant role in solute transport. The trajectories of methanol molecules explore a wider path than the trajectories of ethanol in the upper pathway A; water molecules explore an even wider range of trajectories in the upper pathway A. And there are clearly more methanol and ethanol trajectories in the same amount of simulation time than for 2-propanol or urea. The number of pathways plotted in each panel of Fig. 1 is indicative of the rejection of those molecules in the order 2-propanol > urea > ethanol > methanol > water during the approximately 23 ns duration of the simulation, consistent with previous results [23].

The preferred pathways for both water and solutes indicate that the membrane is inhomogeneous at the nanoscale. Of course, these are not "pores" in the membrane in that the tangled polymer chains of the membrane crisscross the region occupied by these preferred pathways. Instead, these pathways represent regions in the membrane structure where the tangle of the membrane molecular chains is such that the water and smaller solute molecules can pass from one side of the membrane to the other. In fact, this inhomogeneity in the nanoscale membrane structure can undermine solute rejection. All this suggests the potential for improving the water flux and solute rejection by using specific arrangements of polymer chains within the membrane, though the method to accomplish this sort of ordered structure in a practical polymeric membrane would be challenging.

The trajectories can be used to determine a path length, defined as the total length of the fitted curve, as a function of molecule size for water and all solutes passing through the membrane via all trajectories (solid squares) and those that pass through the membrane via pathway A only (open squares), as shown in Fig. 2. The path length is longest for water molecules, because they are small enough to explore a large portion of the volume of the membrane. Path lengths for methanol, ethanol, and urea are similar to one another, but the path length is shorter for 2-propanol, probably because of the relatively large molecular size of 2-propanol limits its movements in the constricted space between polymer chains of the membrane.

The transit time of the molecules, defined as the time from when the molecule enters the membrane to the time the molecule leaves the membrane, is shown for molecules passing through any region of the membrane (solid circles) and for molecules following pathway A (open circles). Urea molecules spend the most time in the membrane, even though the dehydrated molecular size is smaller than ethanol and 2-propanol molecules, and urea molecule path lengths are shorter than those for water, methanol, and ethanol. As we will show later, this is likely a consequence of more hydrogen bonding sites for urea molecules than for alcohol molecules. In most cases, the transit time via pathway A is similar to the transit time via all pathways except for ethanol, which seems to have a preference for pathway A, also evident in Fig. 1(d).

The transit speed of the molecules, defined as the average value of the path length divided by the transit time, is shown in Fig. 3. The transit speed for water is fastest due to its small size, followed by methanol. The transit speed for urea is slower than the other solutes. On the other hand, the solute rejection, defined as the number of solute molecules that pass halfway into the membrane when 1000 water molecules pass through the membrane, is in the order 2-propanol > urea > ethanol > methanol (solid circles in Fig. 3). Thus, although 2-propanol passes through the membrane faster than urea, the rejection of 2-propanol is higher than urea. Clearly, solute rejection does not completely depend on transit speed of the solutes. In fact, the transport of solute molecules through the membrane is a series of events: entering the membrane from the solution; passing through the membrane; and escaping the membrane. For 2-propanol, few solute molecules enter the membrane, due to smaller accessible percolated free volume in the membrane corresponding to the larger solute size [23], resulting in only a few molecules passing through the membrane, even though they pass through the membrane more quickly than urea. Urea, despite being smaller than ethanol, passes more slowly through the membrane, probably due to hydrogen bonding as will be shown later, leading to higher rejection of urea than ethanol.

**3.2. The effect of solute hydration on solute rejection**

It has been suggested that the hydration of ions is important for ion rejection [7, 30]: water molecules are attracted to ions due to the strong electrostatic interaction between an ion as a charged point and water dipoles [31]. As a result, water molecules form a shell surrounding an ion, making it more difficult for the ion to pass through the intermolecular space in an RO membrane. However, the situation for neutral solutes is less clear.

To better understand the impact of the hydration shell on solute rejection, we calculate the average number of water molecules within the first solvation shell of all atoms in the organic solutes when they permeate through the membrane. The same quantities are calculated for $Na^+$ and $Cl^-$ ions but using an artificial initial condition in which the cations and anions were placed on opposite sides of the membrane. This was necessary because otherwise no ions even penetrate the membrane during the limited duration of the simulation. Placing cations and anions on opposite sides of the membrane sets up a situation in which ions are driven through the membrane due to the high electrostatic potential difference allowing the examination of the corresponding hydration shell, which should be unaffected by the mechanism driving the transport. To determine which water molecules to include in the hydration shell, we first calculated the Radial Distribution Function (RDF) of the ion-water oxygen or solute donor/acceptor atom-water oxygen pairs [32] for all solutes when they are outside of the membrane in solution. The first solvation shell is defined as the first peak in the RDF, which is shown in Fig. 4 for all solutes. The magnitude of the peak denotes the relative number of water molecules (or, more specifically, water oxygen atoms) in the solvation shell. We set the cutoff distance for water molecules in the first hydration shell as the radius in the RDF corresponding to the first valley after the first peak. Therefore, the cutoff distance is 3.1 Å for $Na^+$ and 4.1 Å for $Cl^-$, similar to recent results [21], and 3.4 to 3.5 Å for all organic solutes. For subsequent analyses, water molecules within these cutoff distances are assumed to be in the first hydration shell surrounding the solute.

As shown in Fig. 5(a) for organic solutes, the number of surrounding water molecules in the bulk solution (Coordination Number) based on the RDF cutoff distances estimated from Fig. 4 is in the order: methanol < urea ≈ ethanol < 2-propanol. This is the same approximate order as the Van der Waals volume. The 2-propanol solutes lose the most water molecules (about 8) when they enter the membrane from the feed solution; ethanol and urea start with fewer surrounding water molecules in the feed solution and lose fewer (about 6) as they pass through the membrane; and methanol is surrounded with the fewest water molecules in both the feed solution and the membrane (also losing about 6 water molecules). On the other hand, both $Na^+$ and $Cl^-$ start with fewer vicinal water molecules and lose fewer water molecules (1-2 for $Na^+$ and 2-3 for $Cl^-$) than the organic solutes (Fig. 5(b)). This suggests that neither the number of water molecules in the solvation shell nor the number of water molecules in the solvation shell that are shed upon entering the membrane explain the differences in the rejections of ions and organic solutes. We further note that recent results for the coordination number for $Na^+$ and $Cl^-$ ions with equal solute concentrations on both sides of the membrane demonstrate similar reduction in the coordination number at the feed side of the membrane, but also indicate an overall decrease in the coordination number to nearly zero at the permeate side of the membrane [21]. The origin of this result is unclear--one would expect the coordination number to be the same in the bulk solution on both sides of the membrane, especially given the initial condition of equal concentrations of solute on both sides of the membrane. Finally, returning to the results in Fig. 5 for both ionic and organic solutes, it is clear that the dehydrated size alone also does not explain rejection differences, since $Na^+$ ions (1.36 Å in radius) are smaller than even methanol (2.23 Å in radius), and $Cl^-$ ions are similar in size (2.3 Å in radius).

Since it is clear that simple hydration and size effects do not explain the different rejections for organic solutes and $Na^+$ or $Cl^-$ ions, we consider the pair interaction energy for every water-solute pair in the first solvation shell, shown in Fig. 6. The average interaction energy E is calculated by averaging the interaction energy of all individual solute-water pairs in the first solvation shell for molecules or ions, both in solution and within the membrane (since the solute location only affects the number of water molecules, but not the interaction energy). It is evident that the average interaction energy between a neutral organic solute and a water molecule in its vicinity (within the first solvation shell) is an order of magnitude smaller than that between $Na^+$ or $Cl^-$ ions and their water shell (black bars in Fig. 6). In other words, it is easy for the organic solutes to shed their water solvation shell compared to $Na^+$ or $Cl^-$ ions. The interaction energy for $Cl^-$ is smaller than that for $Na^+$, most likely because the Van der Waals size of $Cl^-$ pushes water molecules further away from the ion than for $Na^+$ [33]. Note that the interaction energy for both $Na^+$ and $Cl^-$ are an order of magnitude larger than the thermal energy ($kT$~0.6 kcal mol$^{-1}$, where $k$ is the Boltzmann constant and $T$ is temperature), making the first solvation shell of the ions thermally stable. On the other hand, the interaction energy between an organic solute and a water molecule in its first solvation shell that is not hydrogen bonded to the solute is 0.5 to 1.3 kcal mol$^{-1}$ (Fig. 6), which is comparable to the thermal energy. Therefore, it is thermally unstable for a water molecule to stay in the first solvation shell of organic solutes when it is not hydrogen bonded to the solute, and, thus, it likely plays a weak role organic solute rejection. In Fig. 6 it is also evident that the pair interaction energy for urea is about 60% larger than that for the alcohols. This is associated with more hydrogen bonding sites for urea than alcohol solutes, illustrated in Fig. 7 for urea and ethanol, as we will demonstrate in the next section.

### 3.3. Solute rejection and hydrogen bonds

To further investigate the role of hydrogen bonding on the solute rejection, we determined the number of hydrogen bonds associated with the organic solutes as they permeated through the membrane using a geometrical criterion in which we included bonds where the donor-acceptor distance is within 3.5 Å, consistent with the cut-off distances estimated from the RDF calculation, and the donor-hydrogen-acceptor angle is smaller than 30° [34], as shown in Fig. 6 (red bars). The average pair interaction energy for hydrogen-bonded solute-water pairs of about -5 kcal mol$^{-1}$ is about an order of magnitude larger than the thermal energy for all organic solutes. It is also an order of magnitude larger than the interaction energy between an organic solute and its non-hydrogen-bonded solvation shell water molecules (blue bars in Fig. 6). However, the pair interaction energy for hydrogen bonded water molecules with the organic solutes is substantially less than the pair interaction energy for the solvation shell of both Na$^+$ and Cl$^-$ ions. Thus, it is easier for the organic solutes to shed even their hydrogen bonded water molecules than for Na$^+$ or Cl$^-$ ions to shed water molecules in their solvation shell, accounting for the higher rejection of ions than organic solutes.

We can also compare the average number of hydrogen bonds, $N_h$, for organic solute molecules and water, shown in Fig. 8(a). Urea has about double the number of water hydrogen bonds of any of the alcohols, both in solution and in the membrane. This is clearly a consequence of differences in the number of available sites for hydrogen bonding, evident in the molecular structures for urea and alcohols (Fig. 7). Urea has two amine sites and one carboxyl site to form hydrogen bonds, while the three alcohol solutes have only one carboxyl site for hydrogen bonds. When urea molecules enter the membrane (gray area in Fig. 8(a)) from the bulk solution, they shed more hydrogen bonded water molecules than the other organic solutes, and then they regain a similar number of hydrogen bonds when exiting the membrane into the permeate.

Within the membrane, urea molecules also have more hydrogen bonds with the polymer chains of the membrane than the alcohols as the loss of hydrogen bonds with water is partly compensated by hydrogen bonding with polymer (Fig. 8(b)). The number of solute-water, solute-membrane, and solute-solute hydrogen bonds are shown in Fig. 9, both in water (based on both the feed and permeate sides of the membrane) and within the membrane. In water (outside of the membrane), urea has more solute-water and solute-solute hydrogen bonds than any of the alcohols. Upon entering the membrane, all solutes lose some hydrogen bonds with water. Some of the solute-water hydrogen bonds that are lost are replaced with solute-membrane hydrogen bonds, but urea has nearly double the number of solute-membrane hydrogen bonds while also retaining some solute-solute hydrogen bonds. This is even more evident in the supplementary videos (V1 and V2), frames from which are shown in Fig. 10 (a,b). In this figure and the videos, the molecular structure of the membrane is shown in green with the solute molecules in blue, noting that the sizes of urea and ethanol solutes are quite similar (as indicated in Fig. 2). The hydrogen bonding of ethanol molecules and urea molecules with the membrane polymer chains (not with water molecules or other solute molecules) can be compared as the hydrogen bonds are instantaneously labeled red as they occur. At the instants shown in Fig. 10, it is quite clear that there are many more hydrogen bonds for urea than for ethanol.

In the videos, it is quite evident that the solute-membrane hydrogen bonds are continually breaking and reforming as the water molecules in the membrane buffet the solute molecules and the membrane structure vibrates. However, the videos also indicate that the solute-membrane hydrogen bonds for urea are on average longer in duration than for ethanol because there are more chances for urea to have multiple hydrogen bonds with the membrane than for ethanol based on more hydrogen bonding sites for urea than for ethanol (see Fig. 7). The nature of the hydrogen bonding for methanol and 2-propanol (not shown in Fig. 10 or the videos) is similar to that for ethanol.

The differences in hydrogen bonding explains the higher rejection of urea than ethanol, despite the Van der Waals size of urea being smaller than ethanol: breaking hydrogen bonds with water increases the energy barrier as urea molecules enter the membrane, and breaking hydrogen bonds with the polymer chains increases the energy barrier to escape the "traps" within the membrane. Both make it more difficult for urea molecules to travel across the membrane than ethanol molecules. Note further that previous results [23] as well as trajectories in Fig. 1 suggest that the trajectories of urea molecules within the membrane may be somewhat more widely spread than trajectories of ethanol molecules within the membrane, perhaps qualitatively explained by more opportunities for urea to be hydrogen bonded with polymer chains of the membrane. On the other hand, due to the smaller size of urea molecules than 2-propanol molecules, urea molecules can explore more percolated free volume than 2-propanol molecules [23]. Hence, the rejection of urea does not exceed that of 2-propanol, consistent with experimental results [5].

Thus, the order of organic solute rejection indicated in Fig. 3 is a result of competition between the impact of physical effects associated with solute size membrane pore structure and chemistry effects in terms of the solute-water and solute-membrane interactions. Van der Waals sizes of the solutes, solute hydration, and the number of hydrogen bonds play different roles in different situations. For the alcohol solutes under study, the solute rejection increases with the size of the dehydrated solutes, which is related with reduced percolated free volume for larger solutes. However, the size dependence of solute rejection breaks down for urea, where more hydrogen bonding occurs with water, other urea molecules, and the membrane.

Returning now to the ions, the ion-water pair interaction energy is at least twice as large as the hydrogen bond energy for organic solutes (Fig. 6), and it is less sensitive to orientation than the hydrogen bond interaction [31], making ion-water pairs more stable than the hydrogen bonded pairs. This can be corroborated by the calculation of escape lifetime for solute-water pairs while in solution. We use both a direct calculation of a continuous lifetime, defined as the average time a water molecule is continuously within the solvation shell [35] and a correlation function calculation of an intermittent lifetime, $c(t)$ [36]:

$$c(t) = \frac{\langle h(t)h(0) \rangle}{\langle h(0)^2 \rangle} \tag{1}$$

where $t$ is lag time and $h(t)$ is 1 when a water molecule is in a solvation shell (for organic solutes when the water molecule is in the vicinity of a donor/acceptor atom) and 0 when it is not. The intermittent lifetime is extracted as the characteristic time for the decay of $c(t)$ [36]. Note that the intermittent lifetime includes water molecules that temporarily escape from the solvation shell and return later on. Therefore, the intermittent lifetime is typically longer than the continuous lifetime.

Results for the intermittent and continuous lifetimes are listed in Table 1 for all solutes, and the correlations for the intermittent lifetimes are shown in Fig. 11. Ideally, the decay of the correlation $c(t)$ should be exponential, but this is not always the case [36]. In fact, the decay of the correlation is somewhat different from exponential for the organic solutes, though it is nearly exponential for $Na^+$ and $Cl^-$. Nevertheless, we use the value where $c(t_0) = e^{-1}$ as the arbitrary level to estimate the characteristic intermittent lifetime $t_0$. We further note that the characteristic lifetimes (both intermittent and continuous) are only estimated to ±1 ps due to the 2 ps time resolution of the simulation data that was saved for analysis.

Table 1 shows that the continuous lifetimes are generally shorter than the intermittent lifetimes, as expected. More importantly, both the intermittent and continuous lifetimes for ions are substantially longer than those for organic solutes, confirming the tendency for ionic solutes to retain water molecules in their first solvation shell longer than organic solutes. In addition, comparing Fig. 5(b) to Fig. 8(a), it is evident that while in the membrane, the fraction of water molecules that $Na^+$ and $Cl^-$ ions retain in the first solvation shell is larger than the fraction of hydrogen bonded water molecules that organic solutes retain. Clearly, organic solutes more easily lose water molecules, even those that are hydrogen bonded, from their solvation shell than ions because of the lower energy penalty (see Fig. 6).

## 4. Conclusion

From these molecular dynamics simulations it is evident that the rejection of contaminant solutes does not always increase with dehydrated solute size, consistent with experimental results [5]. The solute rejection is positively correlated with the Van der Waals size of the dehydrated solutes for alcohol solutes due to decreasing accessible free volume with solute size [23]. However, the dependence of solute rejection on the dehydrated size of the solutes breaks down for ions and urea due to chemistry effects. $Na^+$ and $Cl^-$ ions are so strongly bonded with the water solvation shells that their rejections are not correlated with the dehydrated Van der Waals size of the solutes, and hydrogen bonding for urea alters its transport through the membrane compared to alcohols.

Ultimately, it is the free energy penalty that governs solute rejection. Putting entropy aside, the energy penalty is essentially the number of water molecules multiplied by the pair interaction energy. The higher the interaction energy, the more difficult it is to separate water molecules from the solute, leading to fewer water molecules being shed as a solute enters and transits the membrane structure. Nevertheless, the number of water molecules that are shed not only depends on the interaction energy, but also the open space in the molecular structure compared with the hydrated solute size. If the smallest open spaces in the membrane's molecular structure are all larger than the solute size including its hydration shell, then the energy penalty does not matter. However, when the pore size is such that solutes have to shed at least one water molecule to pass through a portion of the membrane molecular structure, the pair interaction energy becomes important. In this sense, the water solvation shell stabilizes the rejection of ions—water molecules in the solvation shell are difficult to remove due to the high pair interaction energy.

The surrounding solvation shell for organic solutes, when not hydrogen bonded, does not play an important role for solute rejection because the pair interaction energy is comparable to the thermal energy at room temperature and the water molecules are easily shed. Even water molecules that are hydrogen bonded to organic solutes are shed much more easily than those from the solvation shell of ions. These differences between ions and organic solutes are evident in the longer escape lifetimes for ions (Table 1) and the larger fraction of water molecule that are retained in the solvation shells of ionic solutes (comparing Fig. 5(b) to Fig. 5(a) and Fig. 8(a)), both of which are results of the higher pair interaction energy for ions than for organic solutes (Fig. 6). Consequently, higher rejection of $Na^+$ and $Cl^-$ than the organic solutes is due to two factors: first, $Na^+$ and $Cl^-$ have more water molecules in their first solvation shell than organic solutes have water molecules that are hydrogen bonded to them; and second, the pair interaction energy for water molecules in an ion's solvation shell is much larger than the pair interaction energy for water molecules that are hydrogen bonded to organic solutes. This makes the water molecules in the ion's solvation shell more resilient to being shed so the solute can squeeze through small open spaces between the membrane's polymer chains, thus improving the rejection of ions compared to organic solutes that are similar in size or even larger. Furthermore, water molecules that are hydrogen bonded to organic solutes are important for rejection, since the pair interaction energy between a solute and hydrogen bonded water molecule is so much larger than the thermal energy. Thus, the higher rejection of urea than ethanol can be attributed to more hydrogen bond sites for urea molecules. This leads to high energy barriers for urea to enter, pass through, and escape from the membrane.

The exploration of more types of solutes, larger simulation systems, other membrane structures, and longer simulation times can be carried out using the framework that we have established here to further establish both size effects in terms of solute size and membrane structure at the molecular level and chemistry effects in terms of solute-water, solute-solute, and solute-membrane interactions.

**5. Acknowledgements**

M. S. thanks Dr. Marco A. A. Corvalan from Northwestern University for useful discussions, as well as Northwestern University High Performance Computing Center for a supercomputing grant. Part of this work also used the Extreme Science and Engineering Discovery Environment (XSEDE) [37], which is supported by National Science Foundation grant number ACI-1053575. The authors gratefully acknowledge funding from the Institute for Sustainability and Energy at Northwestern (ISEN).

**Supplementary materials:**

**Video V1: Hydrogen bonding between ethanol solute molecules and the membrane during filtration.**

**Video V2: Hydrogen bonding between urea solute molecules and the membrane during filtration.**

Table 1. Average intermittent and continuous lifetimes (to ±1 ps) of solvation water molecules for ions and hydrogen bonded water molecules for organic solutes.

| Solute | Intermittent Lifetime (ps) | Continuous Lifetime (ps) |
|---|---|---|
| $Na^+$ | 41 | 25 |
| $Cl^-$ | 20 | 8 |
| Methanol | 3 | 4 |
| Ethanol | 6 | 5 |
| 2-propanol | 7 | 5 |
| Urea | 5 | 4 |

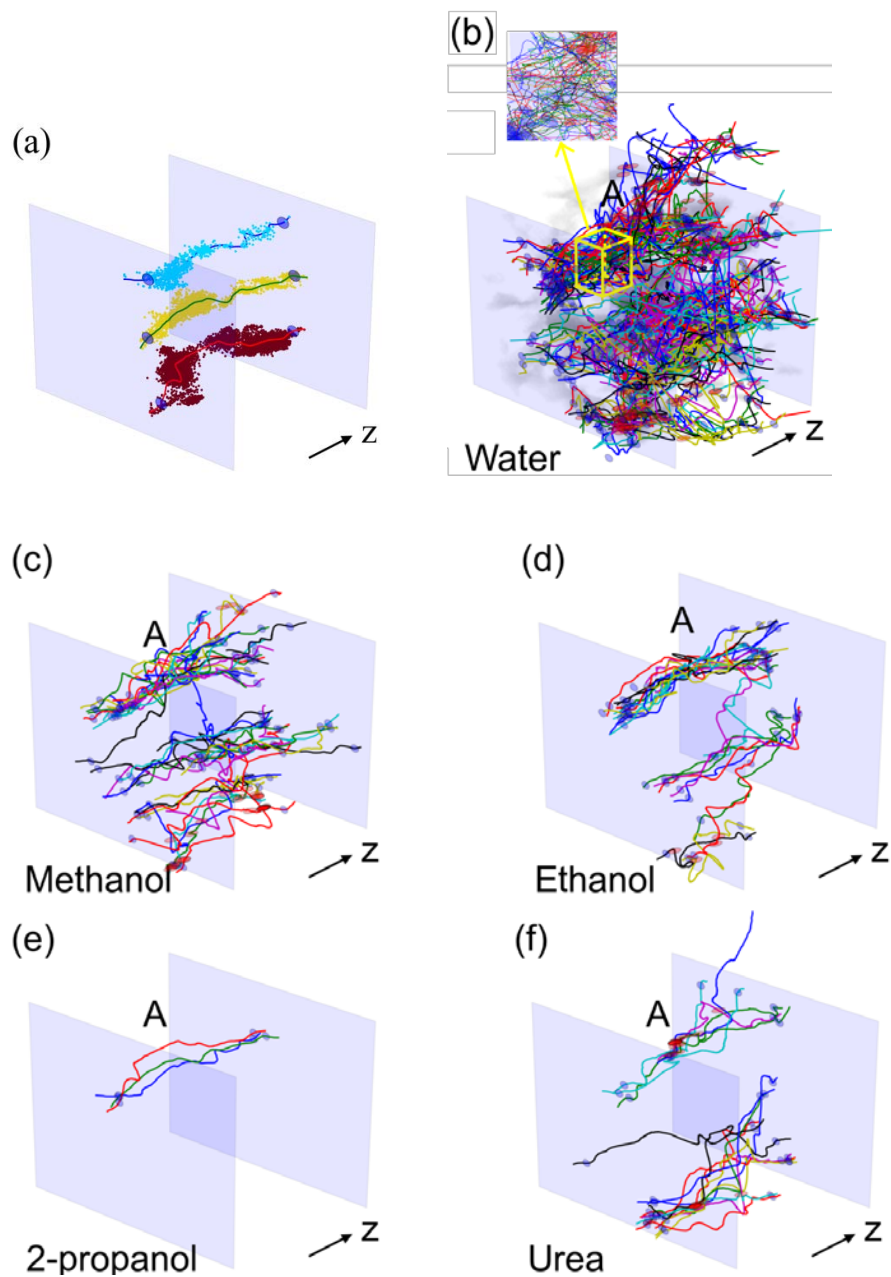

Fig. 1. (a) The positions of 2-propanol solutes as they transit through the membrane and the molecule trajectories fitted using the Savitzky-Golay algorithm. The trajectories are shifted so they do not overlap for clarity. The in-membrane trajectories of (b) water, (c) methanol, (d) ethanol, (e) 2-propanol, and (f) urea molecules. The inset in (b) shows trajectories for water molecules within the region of the yellow cube. The transparent planes represent the boundary of the dense region of the membrane, and small circles on the planes indicate where trajectories intersect the planes. (Color online.)

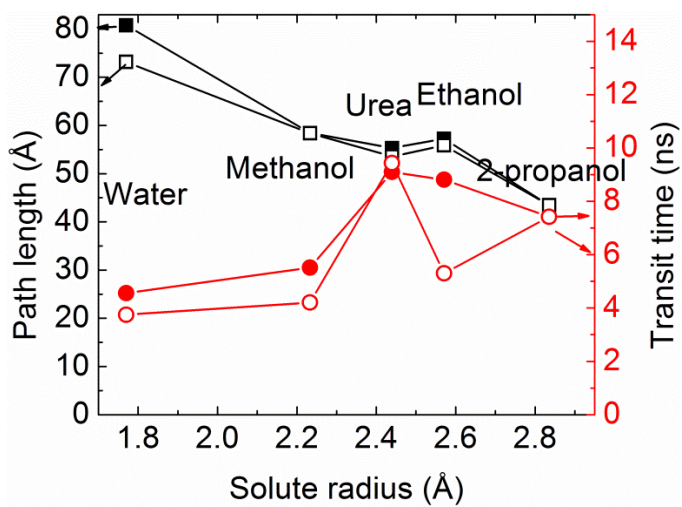

Fig. 2. Average path length (squares) and transit time (circles) for water, methanol, ethanol, 2-propanol and urea through the membrane as a function of solute radius, assuming a spherical shape, for all trajectories (solid symbols) and for trajectories only in region A (open symbols) for molecules that cross the entire membrane. Transit time is the longest for urea, even though the size of a urea molecule is the second smallest solute. (Color online.)

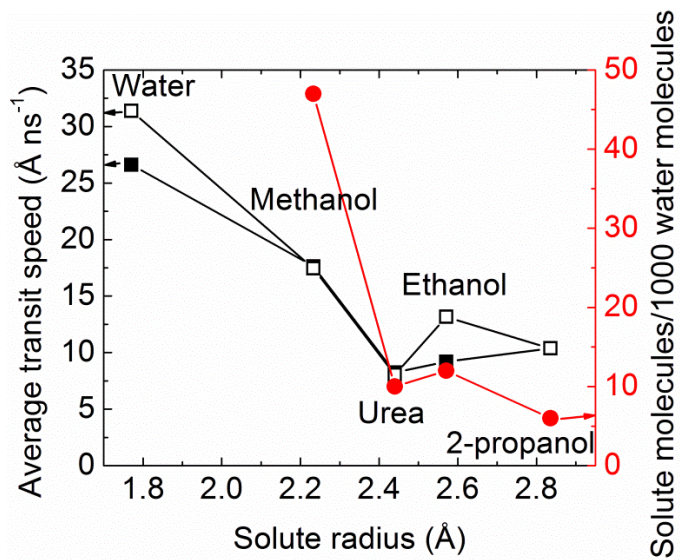

Fig. 3. The number of solute molecules that pass half way through the membrane (solid circles) when 1000 water molecules have passed through the membrane and the average transit speed (squares) along the molecule trajectories in the membrane for water, methanol, ethanol, 2-propanol and urea for all trajectories (solid squares) and for trajectories in region A (open squares). (Color online.)

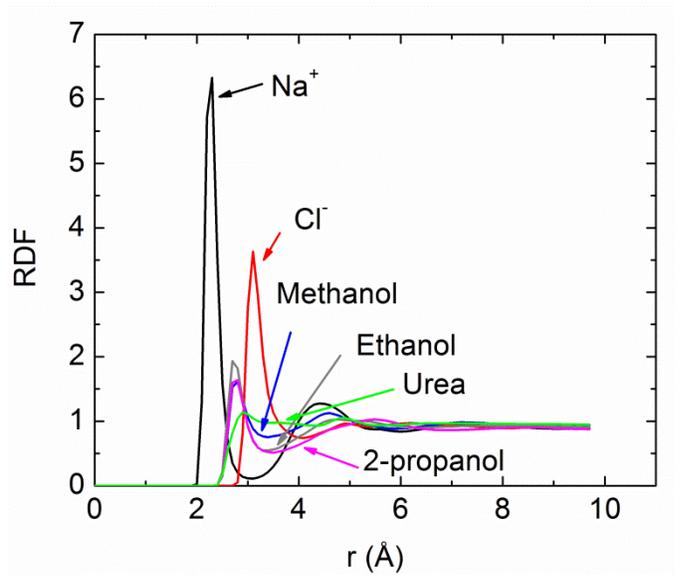

Fig. 4. Radial Distribution Function (RDF) of ion-water oxygen atom and solute donor/acceptor atom-water oxygen atom pairs. (Color online.)

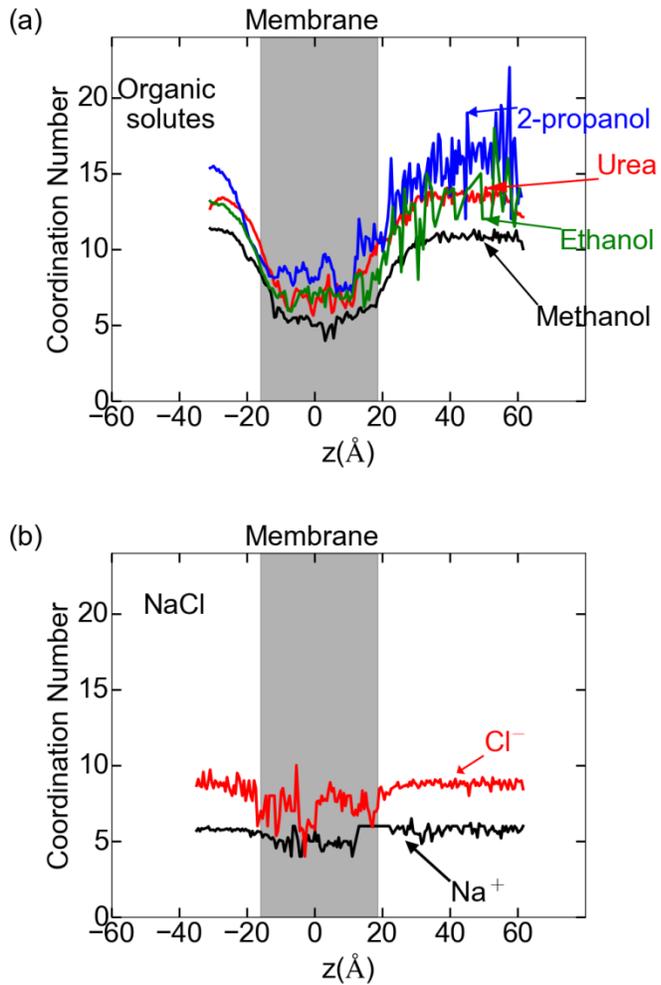

Fig. 5. The average number of water molecules, Coordination Number, within the first solvation shell for (a) the organic solutes and (b) Na$^+$ and Cl$^-$, as a function of the z coordinate, which is through the thickness of the membrane. The gray area represents the membrane region. (Color online.)

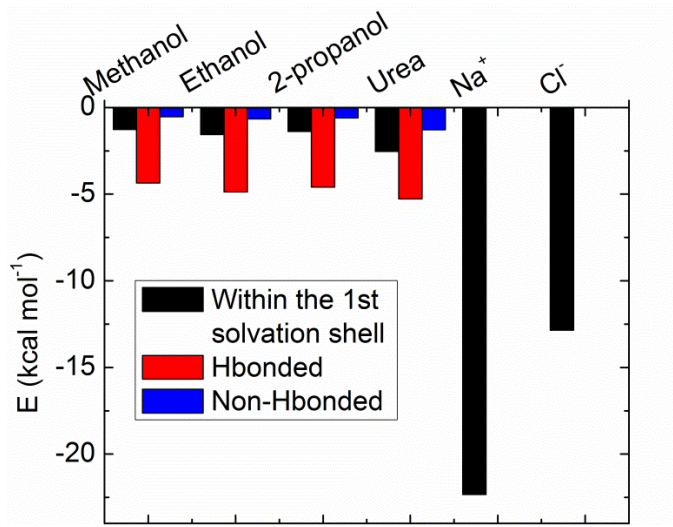

Fig. 6. Average pair interaction energy between a solute and a water molecule within its first solvation shell (black bars), between a solute and a water molecule that is hydrogen-bonded to it (red bars), and between a solute and a non-hydrogen bonded water in its first solvation shell (blue bars). (Color online.)

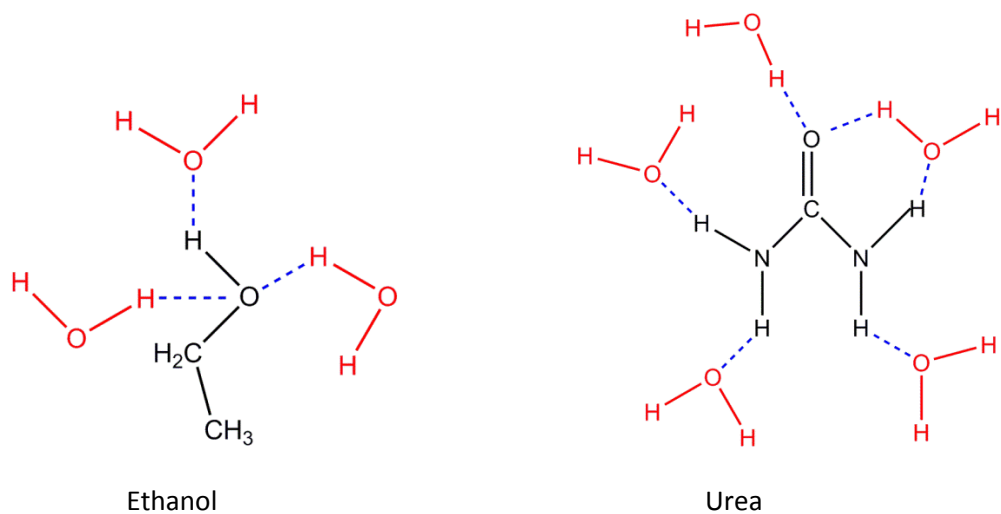

Ethanol                                               Urea

Fig. 7. Illustration of ethanol and urea molecules with the maximum possible hydrogen bonds with water molecules. (Color online.)

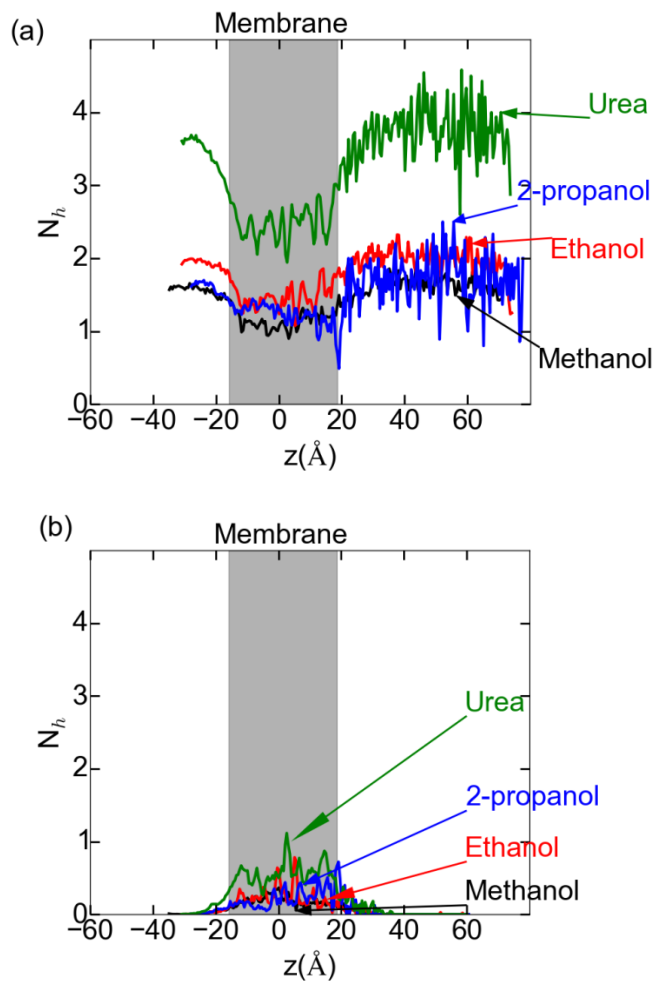

Fig. 8. The average number of hydrogen bonds, $N_h$, (a) between a solute molecule and water molecules and (b) between a solute molecule and membrane, as a function of the z-coordinate. The gray area represents the membrane region. The hydrogen bonds are calculated using the geometrical criteria used in Ref. [34]. (Color online.)

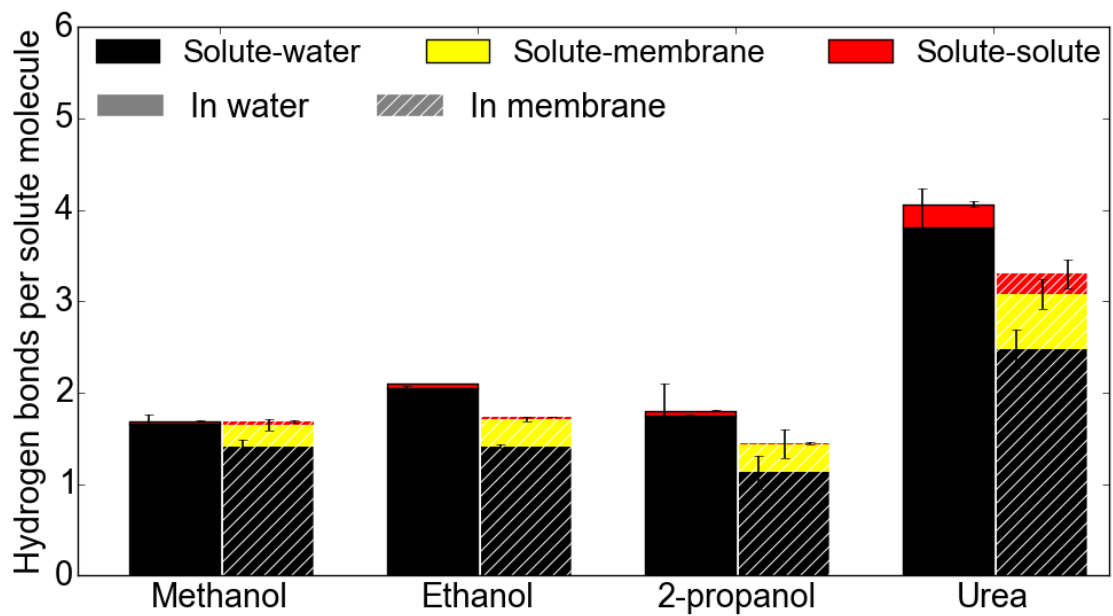

Fig. 9. The average number of solute-water (black), solute-membrane(yellow) and solute-solute (red) hydrogen bonds per solute in the regions of the water solution (without hatches) and the membrane (with hatches). The hydrogen bonds are calculated using the geometrical criteria used in Ref.[34]. (Color online.)

(a)

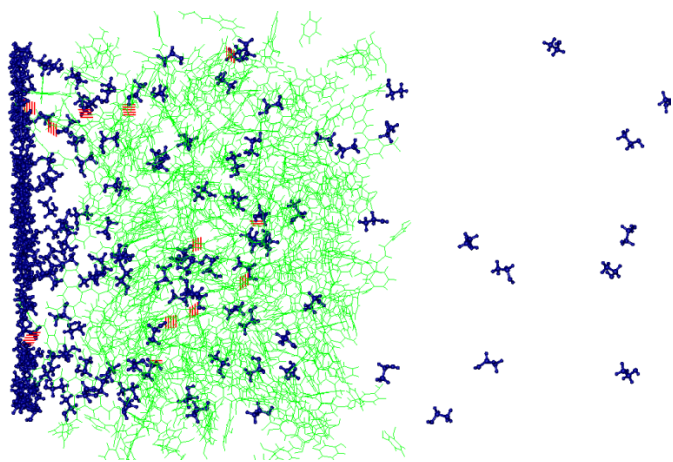

(b)

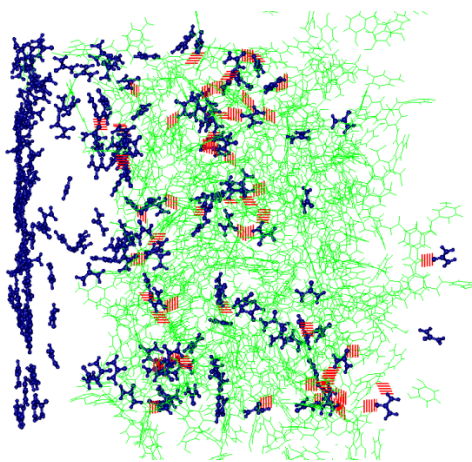

Fig. 10. Snapshots of (a) ethanol and (b) urea solutes (blue molecules) passing through the membrane polymer chain (green). The red marks hydrogen bonds between the solutes and the membrane. (Color online.)

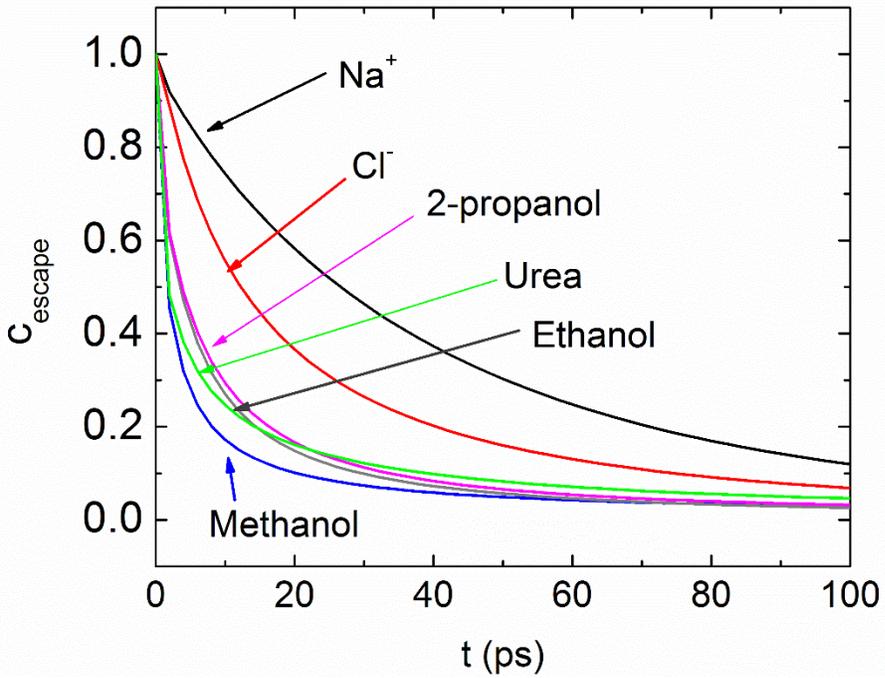

Fig. 11. Correlation $c(t)$ for the escape function for all solutes. (Color online.)